\documentclass{iau}
\usepackage{graphicx}

\def\sss{\scriptscriptstyle}

\def\P{{\sss \!P}}

\title[IAUS290.~~Neutron star properties based on QPOs] 
{Restrictions to Neutron Star Properties Based on Twin-Peak Quasi-Periodic Oscillations}

\author[T\"{o}r\"{o}k et al.]   
{Gabriel T\"{o}r\"{o}k, Pavel Bakala, Eva \v{S}r\'{a}mkov\'{a}, Zden\v{e}k Stuchl\'ik, Martin Urbanec, Kate\v{r}ina Goluchov\'a
}

\affiliation{Institute of Physics, Faculty of Philosophy and Science, Silesian University in Opava,\\
Bezru\v{c}ovo n\'{a}m. 13, CZ-74601 Opava, Czech Republic \\ email: {\tt gabriel.torok@gmail.com }
}

\pubyear{2012}
\volume{290}  
\jname{Feeding compact objects: Accretion on all scales}
\editors{C.M. Zhang, T. Belloni, M. M\'endez \& S.N. Zhang, eds.}

\begin{document}

\maketitle

\begin{abstract}
{
We consider twin-peak quasi-periodic oscillations observed in the accreting low-mass neutron star binaries and explore restrictions to central compact object properties that are implied by various QPO models. For each model and each source, the consideration results in a specific relation between the compact object mass $M$ and the angular-momentum $j$ rather than in their single preferred combination. Moreover, restrictions on the models resulting from observations of  the low-frequency sources are weaker than those in the case of the high-frequency sources.
}
\end{abstract}

\firstsection 
\section{Aims and Scope}

Twin-peak quasi-periodic oscillations (kHz QPOs) appear in the X-ray power-density spectra of several accreting low-mass neutron star (NS) binaries. Observations of the peculiar Z-source Circinus X-1 display unusually low QPO frequencies (\cite[Boutloukos et al., 2006]{bou-etal:2006}). On the contrary, the atoll source 4U~1636-53 displays the  twin-peak QPOs at very high frequencies (e.g., \cite[Barret et al., 2005]{bar-etal:2005}; \cite[Belloni et al., 2007]{bel-etal:2007}). In a serie of works - \cite[T\"or\"ok~et~al.~(2010,~2012)]{tor-etal:2010, tor-etal:2012} and \cite{urb-etal:2010} - we consider these sources and explore restrictions to NS properties that are implied by various QPO models.

\section{Main Findings}

\begin{figure}
\begin{center}
 \includegraphics[width=1\hsize]{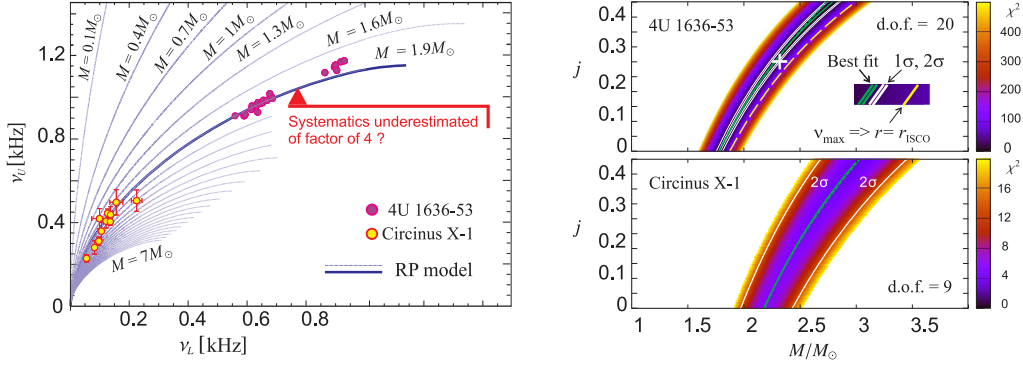}
 \caption{Left: Frequencies predicted by the RP model for $j=0$ vs. data of Circinus X-1 and 4U~1636-53. Right: The quality of the fits for rotating NS. The green line indicates the best $\chi^2$ for a fixed $M$. The white lines indicate the corresponding 1$\sigma$ and 2$\sigma$ confidence levels. The white cross-marker indicates the value found for the RP model by \cite{lin-etal:2011}. The dashed-yellow line indicates a simplified estimate on the upper limits on $M$ and $j$ assuming that the highest observed QPO frequency corresponds to the innermost stable circular orbit (ISCO).}
   \label{figure:1}
\end{center}
\end{figure}

\begin{figure}
\begin{center}
 \includegraphics[width=1\hsize]{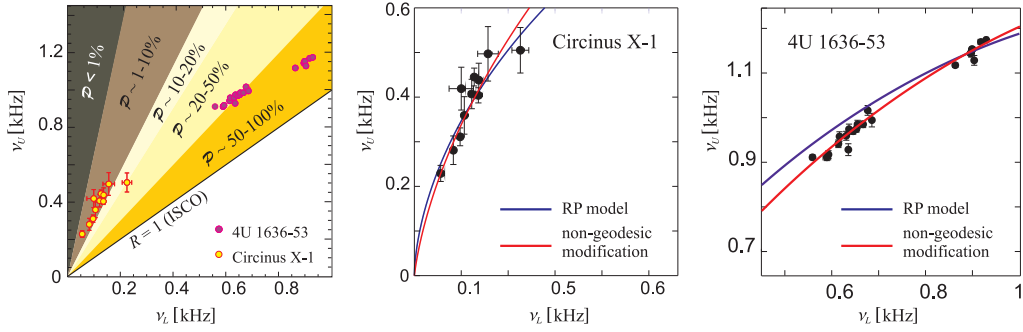}
 \caption{Left: The predictive power of the RP model proportional to the displayed quantity $P$ depends on the source position in the frequency diagram. Moreover, it is related to the ratio between the QPO frequencies rather than to their magnitude. Middle and Right: Geodesic vs. non-geodesic fits of the Circinus X-1 and 4U~1636-53 data. See \cite{tor-etal:2012} for details.}
   \label{figure:2}
\end{center}
\end{figure}

For each twin-peak QPO model and each source, the consideration results in a specific relation between the NS mass $M$ and the angular-momentum $j$ rather than in their single preferred combination.
We also observe some differences in the $\chi^2$ behaviour that represents a dichotomy  between the high- and the low- frequency sources. In general, the low-frequency sources data are matched by the models better than those of the high-frequency sources. Based on the relativistic precession (RP) model introduced by \cite{ste-vie:1999}, we demonstrate that this dichotomy is related to strong variability of the model predictive power across the frequency plane implied by the radial dependence of the characteristic frequencies of orbital motion. As a consequence, restrictions on the models resulting from observations of  the low-frequency sources are weaker than those in the case of the high-frequency sources. These findings are illustrated in Figures~\ref{figure:1}~and~\ref{figure:2}.

For a particular non-geodesic modification of the RP model that we consider in {\cite{tor-etal:2012}}, the data of both classes of sources are well-matched (see Figure~\ref{figure:2} for illustration). The same result is valid for some models assuming non-axisymmetric vertical and radial disc-oscillation modes.

\subsection{Acknowledgements}
The reported work has been supported by the Czech research grants GACR 209/12/P740, GACR 202/09/0772, MSM 4781305903 and the project CZ.1.07/2.3.00/20.0071 - "Synergy" supporting international collaboration of the Institute of Physics at SU Opava. The authors further acknowledge the internal grant of SU Opava, SGS/1/2010.


\end{document}